\documentclass[twocolumn]{aastex631}

\usepackage{amsmath}
\usepackage{enumitem}
\usepackage{rotating}
\usepackage{booktabs}
\usepackage{longtable}
\usepackage{array}
\usepackage{geometry}
 \usepackage{supertabular}
\usepackage{cuted}
\usepackage[utf8]{inputenc}
\DeclareUnicodeCharacter{030C}{\v{c}}

\shorttitle{critical review - interstellar}
\shortauthors{Brown and Borovi\v{c}ka}
\graphicspath{{./}{figures/}}

\newcommand{\degree}{{^{\circ}}}

\begin{document}

\title{On the Proposed Interstellar Origin of the USG 20140108 Fireball}

\author[0000-0001-6130-7039]{Peter G. Brown}
\affiliation{Department of Physics and Astronomy \\
University of Western Ontario \\
London, Ontario, N6A 3K7, Canada}
\affiliation{Institute for Earth and Space Exploration \\
University of Western Ontario \\
London, Ontario, N6A 3K7, Canada}

\author[0000-0002-5569-8982]{Ji\v{r}\'{\i} Borovi\v{c}ka}
\affiliation{Astronomical Institute of the Czech Academy of Sciences, Fri\v{c}ova 298, CZ-25165 Ond\v{r}ejov, Czech Republic}




\begin{abstract}
A critical review of the evidence for the interstellar origin for the USG 20140108 fireball is presented. Examining USG fireball velocities where independent data are available shows the former to have significant (10-15 km/s) uncertainties at large speeds and highly variable radiant accuracy, with average errors in excess of ten degrees. Ablation model fits to the observed lightcurve are possible for normal chondritic impactors only assuming low speeds. To match the high speed and low fragmentation height of the USG 20140108 fireball would require a high density/strength object with low drag and highly aerodynamic shape not made of iron. We suggest the simpliest explanation for the unusual characteristics of USG 20140108 is that the speed, in particular, is substantially overestimated.
\end{abstract}



\section{Introduction}\label{sec:intro} 
In the early 20th century the field of meteor science was driven largely by the question of the interstellar vs. interplanetary nature of most meteors \citep{Hughes1982, Gunn2005}. This controversy developed due to the large fraction of apparently hyperbolic meteors found from visual observations \citep{Opik1950, Almond1950, Williams2004a}. Optical and radar techniques of higher precision than visual measurements conclusively showed by the mid-20th century that the vast majority of apparently interstellar velocities were the result of measurement errors \citep{Lovell1954, Stohl1970, Hajdukova2019}. 

More recent ground-based optical surveys focused on interstellar meteor detection \citep[e.g.][]{Hajdukova1994, Hawkes1997a, Musci2012} have not found convincing evidence ($>$ 3$\sigma$) of any clear hyperbolic population at mm-cm sizes. \citet{Hajdukova2020a, Hajdukova2020b} have analysed optical meteor datasets and find a clear correlation between orbit determination quality and the fraction of meteors appearing to be hyperbolic.

At smaller meteoroid sizes, radar-based measurements have proven more equivocal. \citet{Taylor1994, Baggaley2000} have claimed detection of interstellar meteoroids at tens of micron sizes using the Advanced Meteor Orbit Radar (which ceased operation in 2000). Their most notable result was the claim of an interstellar meteoroid stream emanating from the direction of the debris-sisk star $\beta$-Pictoris (though the interpretation is questioned by \citet{Murray2004}). In contrast, \citet{Weryk2004, Froncisz2020} used the Canadian Meteor Orbit Radar to search for hyperbolic meteoroids at larger sizes (of order several hundred microns) and found only a handful of potential detections out of millions of orbits, which they ascribed to measurement error.

 While ground-based techniques using the Earth's atmosphere as a detector have failed to provide clear evidence for an interstellar meteoroid population at tens of microns - decimeter sizes, spacecraft measurements at small sizes are abundant. The in-situ dust detectors on the Galileo, Cassini and Ulysses spacecraft have unambiguously detected the inflow of interstellar dust at sub-micron sizes in the middle-outer solar system \citep{Grun1993, Kruger2001, Altobelli2003}.

The recent telescopic detection of two interstellar objects in the inner solar system, namely 1I/'Ouamumua and 2I/Borisov \citep{Jewitt2022}, have reinvigorated the search for interstellar meteoroids in the Earth's neighborhood.
\citet{Siraj2022AJ} have recently claimed detection of a meter-sized interstellar meteoroid detected on Jan 8, 2014 based on data from US Government sensors. This is a surprising result given that numerous surveys at smaller meteoroid sizes have searched for and as yet not found convincing evidence for true interstellar meteoroids from ground-based detectors \citep{Hajdukova2019, Froncisz2020, Musci2012}. The claim is impossible to verify conclusively as independent analysis of the original data is not possible since the data originates from classified US Government sensors \citep{Brown2015}. 

However, while specific data for this event are not open for analysis, the general accuracy of US Government sensor observations are better known through comparison with common fireballs detected using other instruments \citep{Devillepoix2019}.
The crux of the interstellar origin claim for USG 20140108 is the measured velocity vector. 

Here we examine the metric data and other information related to this fireball in an effort to better interpret its history and physical properties.

\begin{table}
    \centering
    \label{tab:usgdata}
         \caption{Published data for USG 20140108. The time, location, altitude and speed refer to the point of peak brightness. The three velocity components are given in Earth Centered Fixed coordinates. The total radiated energy represents the estimated electromagnetic radiation emitted during the entire fireball in the silicon bandpass assuming a 6000 K blackbody \citep{Tagliaferri94}.}
    \begin{tabular}{lc}
        \toprule
        Parameter & Value \\
        \midrule
        Date/Time of Peak Brightness & 2014-01-08 17:05:34 \\
        Latitude & 1.3S \\
        Longitude & 147.6E \\
        Altitude (km) & 18.7 \\
        Speed (km/s) & 44.8 \\
        $V_x$ (km/s) & -3.4 \\
        $V_y$ (km/s) & -43.5 \\
        $V_z$ (km/s) & -10.3 \\
        Total Radiated Energy (J) & 3.1$\times$$10^{10}$ \\
        \bottomrule
    \end{tabular}
\end{table}

\section{Available data for USG 20140108}  \label{sec:data}
    \subsection{US Government Sensor Data}
The USG website \footnote{https://cneos.jpl.nasa.gov/fireballs/} is the source of the metric and lightcurve information related to USG 20140108. Table \ref{tab:usgdata} summarizes the primary data from this source.

There are no documented uncertainties for these data - the precision as published is the only constraint on these values. 

Compared to all other events in the USG dataset, USG 20140108 stands out in several respects. Of the 280 fireballs which have speeds and velocity components, it is the 2nd highest reported speed and the 6th lowest altitude, an extremely unusual combination. Of the fireballs in the USG dataset with altitudes of peak brightness below 20 km, all others have speeds half that (or lower) than USG 20140108. Similarly, of all events with speeds above 40 km/s, the next lowest altitude of peak brightness is 46.3 km. Clearly, USG 20140108 is an extreme outlier.

\begin{figure*}[!ht]
    \centering
    \includegraphics[width=.99\textwidth]{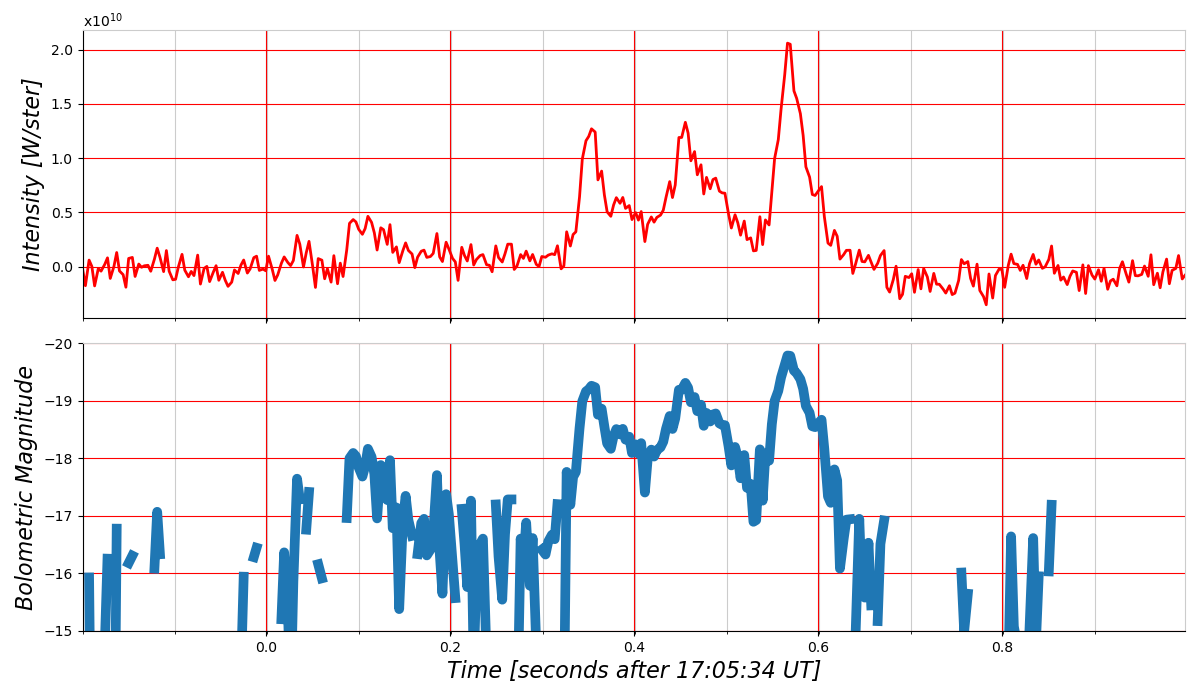}
       \caption{ \label{fig:lc}Light curve for USG 20140108. The top plot shows the radiant intensity as a function of time while the lower plot shows the equivalent bolometric magnitude assuming a 6000K blackbody. The intensity appears to drop below zero in the upper plot due to background subtraction; here the magnitude approaches the limit near -16 to -17.}
    \end{figure*}

Complementing these data is the recently released lightcurve, reporting brightness as a function of time for USG 20140108. This is shown in Figure \ref{fig:lc}. The main feature of the lightcurve is the presence of three equally spaced flares separated by only a few tenths of a second. Using the radiated energy to total energy conversion from \citet{Brown2002}, the estimated total energy for the fireball is 0.11 kT TNT (1 kT = 4.185$\times$10$^{12}$J). 

    \subsection{Infrasound}
    In addition to the optical detection by USG sensors, a search for infrasound (low frequency sound waves with f$<$20 Hz) signals from the fireball was performed across the International Monitoring System \citep{Pichon2019} operated as part of the Comprehensive Test Ban Treaty Organization (CTBTO). 
    
    Acoustic signals from CTBTO infrasound stations within 5000km of the impactor location were examined. Of these, three stations showed signals which could be confidently associated with the fireball based on arrival time and backazimuth association. Table \ref{tab:infra} summarizes the properties of these signals. Analysis was conducted according to the techniques described in \citet{ens2012, Gi2017}. All signals showed celerities consistent with stratospheric ducting. These detections are summarized in detail in Appendix \ref{sec:infra}. 
    
    \begin{table*}
    \centering
    \caption{Infrasound signal characteristics for USG 20140108. Here station is the International Monitoring System designation for each infrasound array, range is the great circle distance from the USG fireball location to the array (in kilometers) and $\Delta_{Az}$ is the difference between the theoretical and observed airwave backazimuth. The delay is the total time from fireball occurrence to signal detection (given as arrival time in UT) at a given array. Cel is the celerity (mean signal travel speed in km/s) while max amp is the maximum amplitude of the Hilbert transformed signal (in Pa) and Max P-P is the maximum Hilbert transformed peak-to-peak amplitude (in Pa). The Period is the period at maximum amplitude using the zero crossing method described in \citet{ReVelle1997, Edwards2006} while T$_{PSD}$ is the period in seconds corresponding to the peak spectral energy from the power spectral density function of the bolide signal following the approach of \citet{ens2012}.}
    \label{tab:infra}
    \begin{tabular}{|l|l|l|l|l|l|l|l|l|l|l|}
    \toprule
    Station & Range & $\Delta_{Az}$& Delay (sec) & Cel (km/s) & Arrival & Max Amp (Pa) & Max P-P (Pa) & Period(sec) & T$_{PSD}$ \\
    \midrule
    I39PW & 1749.7 & 0.9$^\circ$ & 5726.0 & 0.306 & 18:41 & 0.016 $\pm$ 0.005 & 0.029 $\pm$ 0.001 & 1.24 $\pm$ 0.15 &  1.2 \\ \hline
    I07AU & 2525.1 & -2.5$^\circ$ & 8366.0 & 0.302 & 19:25 & 0.04 $\pm$ 0.04 & 0.06 $\pm$ 0.08 & 1.31 $\pm$ 0.17 &  1.2 \\ \hline
    I04AU & 4819.0 & -1$^\circ$ & 16466.0 & 0.293 & 21:40 & 0.02 $\pm$ 0.01 & 0.03 $\pm$ 0.02 & 1.15 $\pm$ 0.3 &  1.3 \\ \hline
    \bottomrule
    \end{tabular}
\end{table*}

    The period and amplitude (with suitable corrections for winds) of these signals can be used to estimate the source energy following the methodology described in \citet{ens2012, Gi2017}. As has been noted previously, period estimates are usually more robust than those derived from amplitude due to the large influence winds can have on the latter \citep{Silber2019}.
    
    We find a best estimate for the source energy of 0.010 $\pm$ 0.001 kT TNT equivalent using the relationship between multi-station period average and fireball energy as given in \citet{ens2012}. The Air Force Technical Applications Center (AFTAC) period-yield relation derived from nuclear tests \citep{ReVelle1997} produces an identical result. Applying doppler wind corrections to these periods lowers the yield approximately 20\%. 

The average yield using wind-corrected infrasound maximum amplitude produces a source energy estimate of 0.04 $\pm$ 0.03 kT TNT. 
    
    While the yield energies from amplitudes and periods show large differences (as expected), taken together these data suggest a source energy a factor of several to as large as an order of magnitude smaller than the USG estimate. Such large differences betweeen USG energy and infrasound energy are atypical, with normal agreement being within a factor of two for events where multi-station infrasound data can be averaged (see Fig 16 in \citet{ens2012}). The infrasound source energy is comparable to the nominal optical energy, implying an unphysically high luminous efficiency approaching 70\%. Such a large apparent luminous efficiency has only been documented for one other fireball, namely the April 23, 2001 Pacific bolide \citep{Brown2002d}.

    This is yet another peculiar aspect of USG 20140108.
  
 \section{Establishing uncertainties of USG through comparison with Ground-based measurements} \label{sec:comp} 

The available metric data for USG 20140108 does not permit a detailed analysis of uncertainties for this particular case. We can, however, gain insight into the general size of USG uncertainties by comparing fireballs simultaneously detected by USG sensors and ground-based instruments. 

\citet{Devillepoix2019} performed such an analysis for USG data available through late 2018. They found that the time of the fireball, the altitude of peak brightness, fireball energy and location were most consistent with ground-based records (within the reported precision of the USG data). The radiant and particularly the speed were least reliable. However, their analysis was largely qualitative and limited to only 10 events. 

In Table \ref{tab:USgcomp} we summarize all 17 USG events where ground-based or other independent measurements are also available as of mid-2023. 

As the main focus of the interstellar nature of USG 20140108 depends on its measured velocity vector we focus on the accuracy of USG measurements of speed and radiant direction.

\begin{figure*}[!ht]
    \centering
    \includegraphics[width=.99\textwidth]{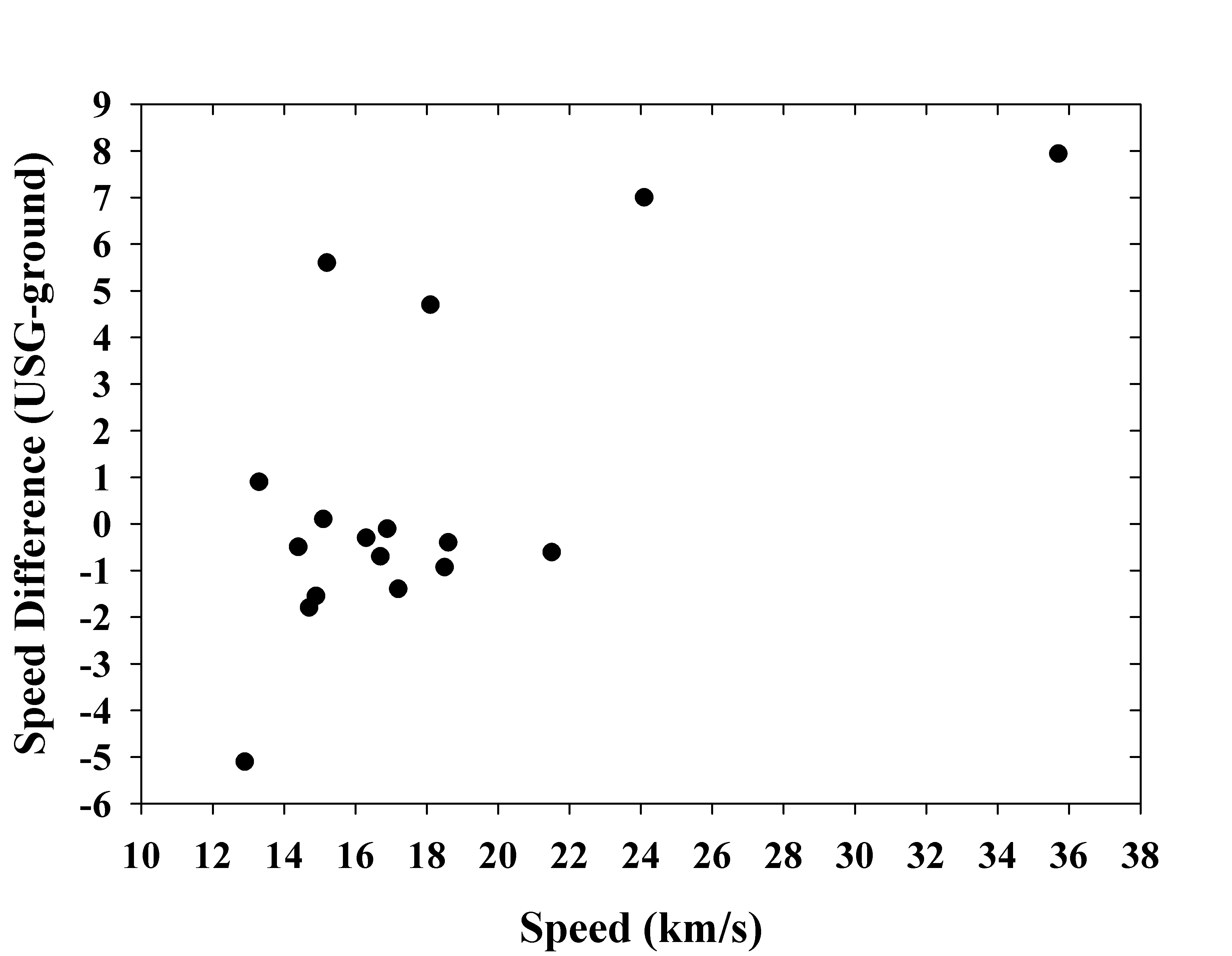}
    \caption{Speed difference (USG-ground-based) as a function of USG reported speed among USG fireballs for which independent speed estimates are available. Data are from Table \ref{tab:USgcomp}.}
    \label{fig:speed-dif}
\end{figure*}

Using the values from Table \ref{tab:USgcomp} we show in Figure \ref{fig:speed-dif} the apparent speed difference between USG speeds and ground-based speeds as a function of speed. We expect a priori that any technique used by USG sensors to sample position as a function of time will necessarily be limited by sample cadence and, all other things being equal, will result in higher uncertainties at higher speeds. Precisely this trend is also seen with ground-based cameras \citep[e.g.][]{vida2020}.

While the number statistics are low, we do see in Figure \ref{fig:speed-dif} a trend that the highest speed USG fireballs show the largest difference compared to ground-based estimates. Moreover, USG tends to overestimate speeds as the event speed increases. 

\begin{figure*}[!ht]
    \centering
    \includegraphics[width=.99\textwidth]{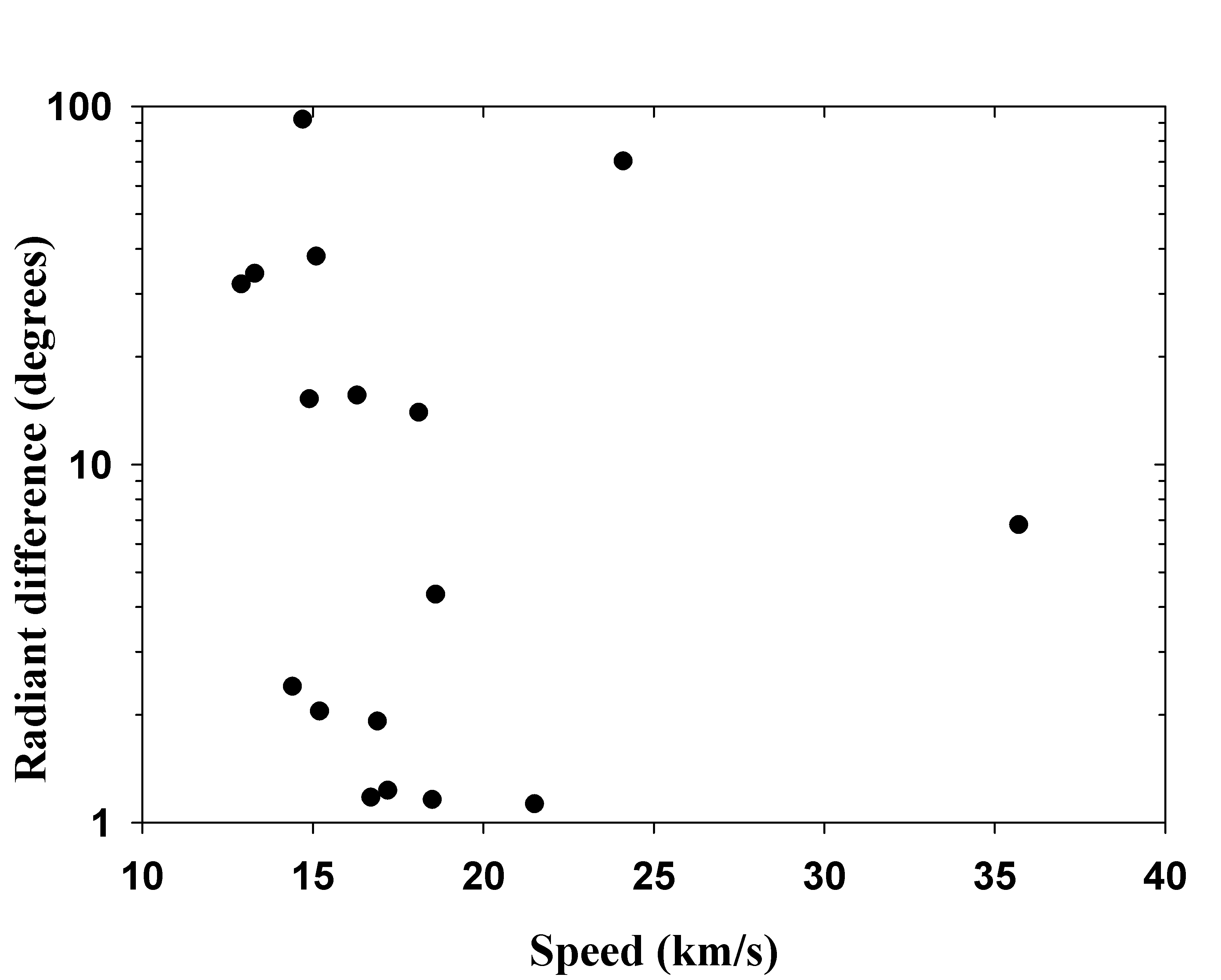}
    \caption{Radiant difference between USG and ground-based data as a function of USG reported speed. Note that the ordinate is a log scale. Data are from Table \ref{tab:USgcomp}.}
    \label{fig:rad-diff}
\end{figure*}

A similar comparison of the radiant difference between USG and ground based data is shown in Figure \ref{fig:rad-diff}. Here no clear trend with speed is apparent, though the uncertainty distribution is bimodal: a significant minority have differences of order a couple of degrees or less (low uncertainty population) while the remainder (high uncertainty population) have differences more than the median which is 10 degrees.  The low uncertainty events have average radiant errors of less than 2 degrees, while the high uncertainty population have a mean and median error of just over 30 degrees. 

Finally, we note that the absolute height difference between the USG peak brightness location and that from ground-based instruments is only 3 $\pm$ 2.6 km and shows no obvious trend with speed. 

One other USG event provides insight into uncertainties, namely the Bering Sea bolide of Dec 18, 2018 . The radiant provided by USG was found by \citet{Borovicka2020bering} to be 13$\pm$9$^{\circ}$ different than that found from triangulation using several satellite views of the dust cloud. While no independent speed estimate was available, we note that the original speed of 32 km/s posted on the CNEOS site was later amended to 13.6 km/s, suggesting that factor of two uncertainties in speeds are possible, particularly at higher apparent speeds. 

These results agree qualitatively with those of \citet{Devillepoix2019}, with radiant and speed being much less secure than the height of peak brightness. What does this relationship suggest about the likely uncertainty in the speed and radiant for USG 20140108?

As the uncertainty in the speed correlates positively with speed, a USG speed measurement of 45 km/s could be expected to be 10-15 km/s above the true speed through simple linear regression extrapolation of Figure \ref{fig:speed-dif}. The radiant difference is harder to establish, but at such high speeds it is difficult to imagine the direction vector would be in the low uncertainty group. Assuming association in the high uncertainty group, a plausible radiant error would be of order 30 degrees based on Figure \ref{fig:speed-dif}.

Thus, examination of USG metric data where ground-truth measurements from other instruments are available suggests plausible errors for USG 20140108 are at least 10-15 km/s in speed and $\approx$ 30$\degree$ in radiant direction.

\begin{table*}[t]
\centering
\caption{Comparison of speed, radiants and height of peak brightness for fireballs recorded in common by USG sensors and other techniques. For fireballs producing recovered meteorites, the meteorite name is given; for fireballs not producing meteorites, the name used in the associated reference is used. The first line contains the ground-based data and the second line USG data. The date is given as year-month-day. The columns $\phi$, $\theta$ give the azimuth and zenith angle of the apparent local radiant while $\Delta _{Rad}$ is the angular difference between radiants. The speed is given at the height of peak brightness, and $\Delta_V$ (km/s) is the difference between speeds. Similarly, height (km) here refers to the height of peak brightness and $\Delta$H (km) is the difference in peak brightness height between USG and the other techniques. Also shown are the type of recovered meteorites (eg How - Howardites, Euc - Eucrites ) as well as the reference where independent metric data used for comparison to USG estimates were extracted. For Almahata Sitta, 2019MO and 2022 EB5 the meteoroid was detected prior to impact and the orbit used to compute the radiant of the fireball. References are: 1. \citet{Jenniskens2009}, 2. \citet{Milley2010}, 3. \citet{Borovicka2013a}, 4. \citet{Borovicka2013}, 5. \citet{Devillepoix2019}, 6. \citet{Borovicka2017}, 7. \citet{Unsalan2019}, 8. \citet{Hildebrand2018}, 9. \citet{Jenniskens2021}, 10. \citet{Kartashova2020}, 11. \citet{CeballosIzquierdo2021},  12. \citet{2019MO}, 13. \citet{Borovicka2021a}, 14. \citet{Froslunda}, 15. \citet{Vida2021}, 16. \citet{2022EB5}}
\label{tab:USgcomp}
\begin{tabular*}{\textwidth}{|p{2.0cm}|p{1.5cm}|p{1.0cm}|p{0.5cm}|p{0.7cm}|p{1.0cm}|p{1.0cm}|p{1.0cm}|p{1.0cm}|p{1.0cm}|p{0.4cm}|}
\hline
\textbf{Fireball/ Meteorite name} & \textbf{Date} & \textbf{$\phi^\circ$ } & \textbf{$\theta^\circ$} & \textbf{$\Delta _{Rad}$} & \textbf{V (km/s)} & \textbf{$\Delta_V$ (km/s)} & \textbf{Ht (km)} & \textbf{$\Delta$H (km)} & \textbf{Met Type} & \textbf{Ref} \\ \hline
       Almahata Sitta & 20081008 & 281.7 & 70 & 34.2 & 12.4 & ~ & 37 & ~ & Euc & 1 \\ \hline
~ & ~ & 249.5 & 83.8 & ~ & 13.3 & 0.9 & 38.9 & 1.9 & ~ & ~ \\ \hline
Buzzard Coulee & 20081121 & 348.6 & 22.8 & 31.9 & 18 & ~ & 31 & ~ & H4 & 2 \\ \hline
~ & ~ & 330.6 & 53 & ~ & 12.9 & -5.1 & 28.2 & -2.8 & ~ & ~ \\ \hline
Kosice & 20100228 & 297.1 & 56.3 & 38.1 & 15 & ~ & 36 & ~ & H5 & 3 \\ \hline
~ & ~ & 257.6 & 27.1 & ~ & 15.1 & 0.1 & 37 & 1 & ~ & ~ \\ \hline
Chelyabinsk & 20130215 & 103.5 & 71.5 & 4.3 & 19 & ~ & 30 & ~ & LL5 & 4 \\ \hline
~ & ~ & 99.9 & 74.1 & ~ & 18.6 & -0.4 & 23 & -7 & ~ & ~ \\ \hline
Kalabity & 20150102 & 346.4 & 70 & 14 & 13.4 & ~ & 40.2 & ~ & ~ & 5 \\ \hline
~ & ~ & 331.7 & 71.6 & ~ & 18.1 & 4.7 & 38.1 & -2.1 & ~ & ~ \\ \hline
Romania & 20150107 & 232.2 & 47 & 6.8 & 27.8 & ~ & 42.8 & ~ & ~ & 6 \\ \hline
~ & ~ & 222.9 & 46.4 & ~ & 35.7 & 7.9 & 45.5 & 2.7 & ~ & ~ \\ \hline
Sariçiçek & 20150922 & 319.6 & 46.2 & 70.4 & 17.1 & ~ & 36.5 & ~ & How & 7 \\ \hline
~ & ~ & 48.8 & 61.8 & ~ & 24.1 & 7 & 39.8 & 3.3 & ~ & ~ \\ \hline
Baird Bay & 20170630 & 6.7 & 17.2 & 2 & 9.6 & ~ & 25.7 & ~ & ~ & 5 \\ \hline
~ & ~ & 11.6 & 18.6 & ~ & 15.2 & 5.6 & 20 & -5.7 & ~ & ~ \\ \hline
Crawford Bay & 20170905 & 177.6 & 53.8 & 92 & 16.5 & ~ & 27 & ~ & ~ & 8 \\ \hline
~ & ~ & 280.1 & 76.7 & ~ & 14.7 & -1.8 & 36 & 9 & ~ & ~ \\ \hline
Motopi Pan & 20180602 & 95.2 & 64.9 & 1.9 & 17 & ~ & 27.8 & ~ & How & 9 \\ \hline
~ & ~ & 95.9 & 66.7 & ~ & 16.9 & -0.1 & 28.7 & 0.9 & ~ & ~ \\ \hline
Ozerki & 20180621 & 238 & 12.4 & 2.4 & 14.9 & ~ & 27.2 & ~ & L6 & 10 \\ \hline
~ & ~ & 237.5 & 10 & ~ & 14.4 & -0.5 & 27.2 & 0.0 & ~ & ~ \\ \hline
Vinales & 20190201 & 171.8 & 39.5 & 15.6 & 16.6 & ~ & 22 & ~ & L6 & 11 \\ \hline
~ & ~ & 176.1 & 54.8 & ~ & 16.3 & -0.3 & 23.7 & 1.7 & ~ & ~ \\ \hline
2019MO & 20190622 & 109.7 & 62.1 & 15.2 & 16.5 & ~ & ~ & ~ & ~ & 12 \\ \hline
~ & ~ & 117.9 & 48.4 & ~ & 14.9 & -1.6 & 25 & ~ & ~ & ~ \\ \hline
Flensburg & 20190912 & 188.1 & 65.3 & 1.2 & 19.4 & ~ & 40.7 & ~ & C1-ung & 13 \\ \hline
~ & ~ & 188.5 & 66.4 & ~ & 18.5 & -0.9 & 42 & 1.3 & ~ & ~ \\ \hline
Froslunda & 20201107 & 243.6 & 17.9 & 1.2 & 17.4 & ~ & ~ & ~ & Iron & 14 \\ \hline
~ & ~ & 242.2 & 16.8 & ~ & 16.7 & -0.7 & 22.3 & ~ & ~ & ~ \\ \hline
Novo Mesto & 20200228 & 156.5 & 42.1 & 1.1 & 22.1 & ~ & 35 & ~ & L5 & 15 \\ \hline
~ & ~ & 155 & 42.7 & ~ & 21.5 & -0.6 & 34.5 & -0.5 & ~ & ~ \\ \hline
2022 EB5 & 20220311 & 129.3 & 33.4 & 1.2 & 18.6 & ~ & ~ & ~ & ~ & 16 \\ \hline
~ & ~ & 129.8 & 32.2 & ~ & 17.2 & -1.4 & 33.3 & ~ & ~ & ~ \\ \hline
\end{tabular*}
\end{table*}

\begin{table*}
\caption{Parameters of light curve models}
\label{table:models}
\begin{tabular}{lllllllllllll} \hline
 Model &  Initial & Density & $\Gamma A$ & Lumin. &Initial  & \multicolumn{3}{c}{First Flare} && \multicolumn{3}{c}{Last Flare}\\   \cline{7-9} \cline{11-13}
 & Speed &&&Efficien.&Mass& Height & Speed & Press. && Height & Speed & Press. \\
 & km s$^{-1}$ & kg m$^{-3}$ &&&kg& km & km s$^{-1}$ & MPa && km & km s$^{-1}$ & MPa\\
 \hline
 12  & 12 & 3400 & 0.6 & normal & 45000 & 20.9 & 11.0 & 10&& 18.7 & 10.5 & 13 \\ 
 16  & 16 & 3400 & 0.6 & normal & 15000 & 21.5 & 14.2 & 15&& 18.7 & 13.2 & 21 \\ 
 20  & 20 & 3400 & 0.6 & normal & 6800  & 22.0 & 17.4 & 20&& 18.7 & 15.4 & 28 \\ 
 24  & 24 & 3400 & 0.6 & normal & 4600  & 22.6 & 20.7 & 26&& 18.7 & 17.6 & 37 \\  
 30  & 30 & 3400 & 0.6 & normal & 3400  & 23.4 & 25.8 & 35&& 18.7 & 20.6 & 50 \\ 
 30-lt  & 30 & 3400 & 0.6 & constant & 4100  & 23.5 & 26.1 & 35&& 18.7 & 21.0 & 53 \\  
 45-lt  & 45 & 3400 & 0.6 & constant & 5700  & -- & -- &--&& 18.7 & 29.5 & 100 \\  
 45-lt-hd  & 45 & 8000 & 0.6 & constant & 2100  & -- & -- & --&& 18.7 & 33 & 130 \\    
 45-exotic & 45 & 8000 & 0.2 & constant & 800  & 27.2 & 44 & 54&& 18.7 & 40 & 190 \\    
  \hline
\end{tabular} 
\end{table*}

\section{Ablation and Fragmentation Modelling} \label{sec:model} 

As another independent check on the properties of USG 20140108 we have tried to reproduce the USG reported light curve with the semi-empirical fragmentation
model of \citet{Borovicka-model}. The detected part of the light curve consists of one faint flare
followed after 0.25~s by the series of three bright flares separated mutually by about 0.1~s (see Figure \ref{fig:lc}). 
The last (fourth) flare was the brightest. According to USG, the maximum brightness point occurred 
at a height of 18.7~km. From the previous section, we expect this height to be accurate to of order 3 km.

A series of models was created by varying the initial meteoroid speed. The height of the fourth flare
was set to 18.7~km in all models and the trajectory slope to the vertical line (i.e.\ the zenith distance of the
radiant) was kept at the USG reported value of $63.2\degree$. 
The assumed mass- and speed-dependence 
of the luminous efficiency ($\tau$) was similar to that \citet{Borovicka-model} but was scaled down two times
since the USG absolute magnitudes are computed under the assumption that 3000~W source is needed to produce a zero-magnitude meteor while the model uses 1500~W. Moreover, $\tau$ of small
fragments was increased, which helped to reproduce the high amplitude of the flares. While \citet{Borovicka-model} used
$\tau= 5$\% for large ($\gg 1$ kg) and 2.5\% for small ($\ll 1$ kg) meteoroids at the speed of 15 km s$^{-1}$, 
we used 2.5\% and 2\% (instead of 1.25\%), respectively.

\begin{figure*}[!ht]
    \centering
    \includegraphics[width=0.99\textwidth]{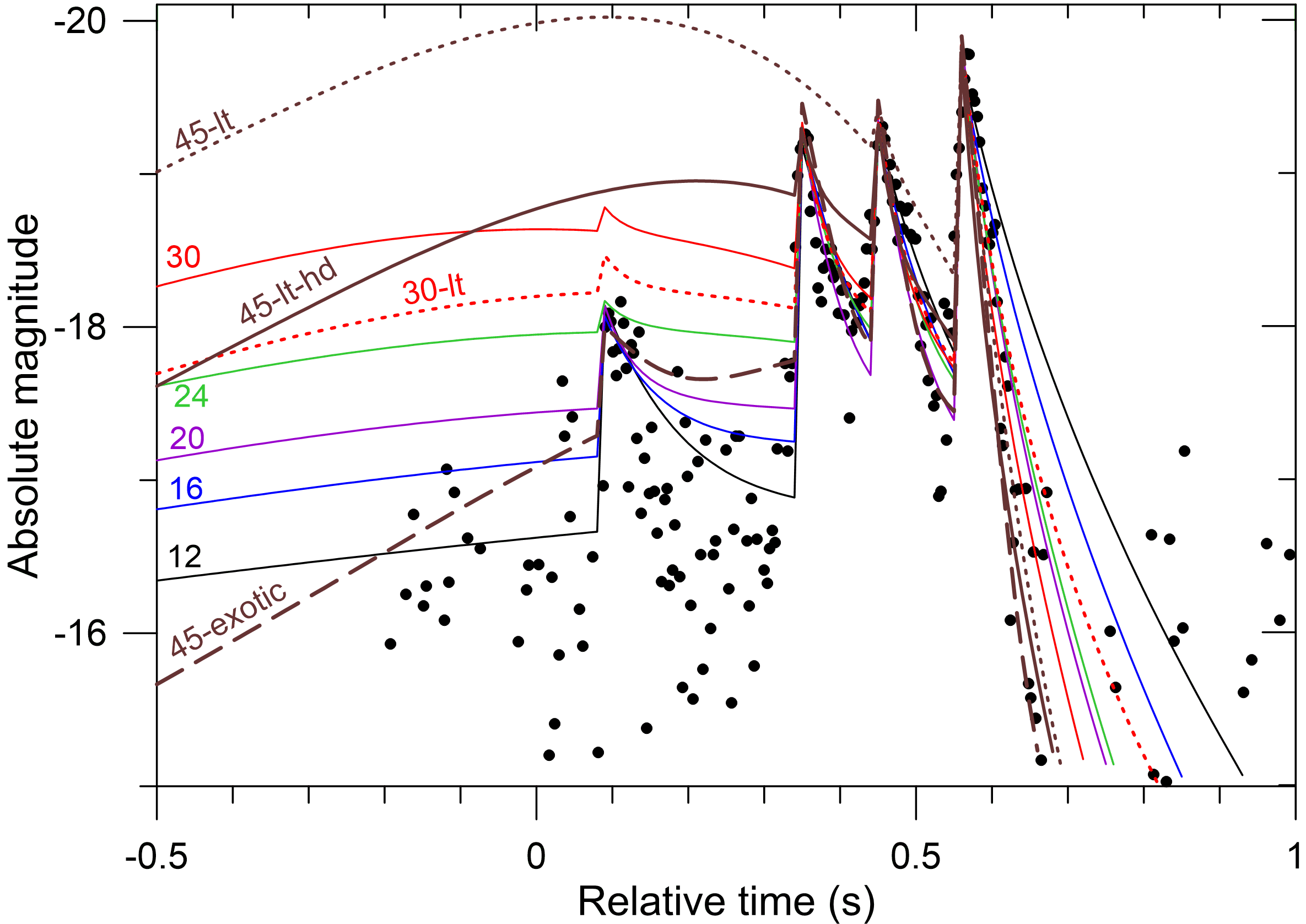}
    \caption{Models of the light curve for various speeds (colored curves) compared with the USG data
    (black dots). See text for details.}
    \label{fig:models}
\end{figure*}

The solid lines in Fig.~\ref{fig:models} shows the computed light curves for five nominal models with
initial speeds between 12 and 30 km s$^{-1}$. These are compared with the USG measurements. The
three main flares could be fitted similarly well in all cases. They were modelled 
by sudden release of small fragments in the mass range 0.1 -- 10 kg from the main meteoroid. 
Nevertheless, there are significant differences between the models in the earlier parts of the light curve, 
namely between the first and the second flare and before the first flare. The models with low initial speeds, 12 and 16 km s$^{-1}$, are consistent with the scattered
data points but for higher speeds the expected light curve becomes too bright. At 30 km s$^{-1}$, the
steady brightness before the second flare is already much higher than the actual brightness 
reached in the first flare.

The parameters of the models are given in Table~\ref{table:models}. The meteoroid density was assumed 
to correspond to typical stony meteorites. The product of the drag and shape coefficients, $\Gamma A$, 
corresponds approximately to spherical meteoroids. The ablation coefficient 0.005 s$^2$km$^{-2}$,
which worked well in fitting fireball data for which meteorites have also been recovered in many instances \citep{Borovicka-model}, was used in all cases. The initial
meteoroid mass was adjusted so that all flares could be produced. As the light curve is a proxy for energy deposition, assuming a constant initial kinetic energy, the lower the speed the higher is the
mass needed to produce a fireball of the desired brightness. 

\clearpage
We have tried to change some of the parameters to be able to reproduce the amplitude of the main flares in the light curve at differing speeds. We note that the luminous efficiency at higher speeds is uncertain.
\citet{Borovicka-model}, following earlier work of Ceplecha \citep[e.g.][]{Ceplecha2005}, assumed
that $\tau$ is increasing linearly with speed above 25 km s$^{-1}$. We have relaxed that assumption
and set $\tau$ to 2.5\% for all speeds and masses.
The corresponding light curve for 30 km s$^{-1}$, designated 30-lt, was then closer to the observations but still
too bright. The discrepancy was much higher for 45 km s$^{-1}$, where the brightness was well above the
first two flares (see model 45-lt in Fig.~\ref{fig:models}). The fragment masses in that model needed 
to be set to 1 -- 10 kg to fit the last flare. 

Next we investigated the case where the meteoroid density was increased to
8000 kg $m^{-3}$, corresponding approximately to metallic meteorites. Still, the early part of the light 
curve was too bright for 45 km s$^{-1}$ (model 45-lt-hd). To be able to produce the first flare 
with such a high initial speed, we had to decrease $\Gamma A$ to 0.2. This would correspond to a very aerodynamic body
with low drag and low cross-section relative to the mass. We will call this model `exotic' henceforth. But even the exotic model
had problems in reproducing the light curve well. The brightness was too high between the first and 
the second flare (see Fig.~\ref{fig:models}). It was possible to remove these discrepancies only by additional decreasing of the ablation coefficient to 0.002 s$^2$km$^{-2}$. The fragment masses in this model were 0.1~kg.

Table~\ref{table:models} also contains the height, speed, and dynamic pressure at the time of
the first and the last flare. Dynamic pressure, computed as $p=\rho v^2$, where $\rho$ is atmospheric
density and $v$ is speed, is an approximate measure of meteoroid strength. 
It is commonly assumed that fragmentation occurs when dynamic pressure exceeds the tensile or shear strength of the meteoroid
\citep{Popova2011,Robertson2017}. In ordinary fireballs studied by \citet{Borovicka-model}, $p$ did
not exceed 5 MPa. The large Chelyabinsk impactor fragmented heavily under 1 -- 5 MPa and only the
strongest daughter fragment reached 18 MPa \citep{Borovicka2013}. 

The data compilation of \citet{Pohl2020}
showed that strengths of meteorites vary widely. Tensile strengths of stony non-carbonaceous meteorites
were measured in the range 2 -- 80 MPa, with the most typical value about 40 MPa. For iron meteorites
it was 40 -- 500 MPa with the mean value being 340 MPa. The mean values of compressive strengths are about 250 MPa, 
and 400 MPa for stones and irons, respectively. Data for shear strengths are not available.
It is evident, that atmospheric fragmentation occurs at lower pressures than the meteorite tensile strength,
at least for stones (atmospheric fragmentation of large iron meteoroids has not been studied as yet). This behaviour is usually explained
as being due to the presence of cracks in incoming meteoroids \citep{Borovicka-model}.

Dynamic pressures for the USG fireball were large in all cases. Only for the lowest speeds below
16 km s$^{-1}$, are they comparable or lower than the maximum dynamic pressure reached in Chelyabinsk. Of course,
we cannot exclude that the meteoroid did not contain any cracks and the fragmentation occurred when
the material strength was reached, a rare but not unheard of situation as exemplified by the formation of the Carancas crater \citep{Brown2008a, Borovicka2008}. If the material was similar to iron, the dynamic pressure even for the highest speed
(190 MPa) is consistent with the material properties. However, it would be strange in that case
that the meteoroid fragmented repeatedly and into large fragments and not into constituent grains
(which would lead to a single terminal flare). 

Moreover, the exotic model is not appropriate for iron
since it uses a low ablation coefficient 0.005 s$^2$km$^{-2}$. Since the melting temperature of iron is lower than
that of stone, the ablation coefficient should be at least several times higher. Under that condition, a 
meteoroid of speed of 45 km s$^{-1}$ would not be able to penetrate to the height of 18.7~km unless much larger
than modelled (and orders of magntitude more energy than indicated by the infrasound records). The fireball would also be much brighter than observed. The exotic model can therefore be applied
only to a hypothetical material of high density and low ablation ability (and body with highly aerodynamic shape).

In summary, lightcurve modelling using the measured speed of USG 20140108 can only reproduce the observed lightcurve/flares for an extremely unusual (high density, low ablation coefficient) object with extremely low drag. In contrast, assuming a lower speed allows reasonable matches to the observed lightcurve for non-exotic entry model parameters, though with a relatively high strength compared to other fireballs, but within the range of stony meteorite tensile strengths.

\section{Discussion}\label{sec:dis} 
In addition to USG 20140108, the dataset of all USG fireballs shows five other fireballs with hyperbolic orbits. The properties of all USG hyperbolic fireballs are summarized in Table \ref{tab:USG_prior}.

To illustrate the speed and radiant uncertainties which can potentially move these events from unbound to bound orbits we plot each event in Figure \ref{fig:hyper}, following the approach of \citet{kresak1976, Hajdukova2019}. Here the elongation of the apparent radiant from the apex of the Earth's way (the direction vector of the Earth's velocity) is shown as a function of the entry speed. The blue line demarcates bound orbits (to the left of the line) from unbound orbits (to the right of the line).

As shown by the figure, most of the USG unbound orbits become bound presuming an overestimate in speed of order a few km/s even if the radiants are perfectly known. This is well within expected uncertainties shown earlier. Two of the events (at 30 and 35 km/s) are closer to 10 km/s over the unbound limit for fixed radiants. However, for these events, adopting an uncertainty in the radiant of only 10-20 degrees result in bound orbits for much smaller velocity overestimates, which is well within expected USG radiant accuracy established earlier. 

The USG 20140108 event (represented by the red cross) is again the most extreme of the unbounded group. It would require an overestimate of more than 20 km/s presuming an accurate radiant to move from bound to unbound. This is not much higher than our extrapolated speed error estimate at such high speeds. However, when allowing for radiant uncertainty as well, an error of 20-30 degrees would translate into a required speed reduction of only 10-15 km/s which together could push the orbit to be bound. These values are compatible with errors of similar magnitude  found in our earlier comparison between USG fireballs also recorded by other instruments. Given the high speed of USG 20140108 compared to our calibration dataset, the required uncertainties to bring the event to a bound orbit are, in our view, very reasonable.

Our lightcurve modelling assuming a chondritic object agrees with the observed lightcurve for cases where speeds are much lower than the nominally measured value of 45 km/s. In particular, we cannot reconcile the apparent lack of light production before the observed flares for an object travelling at even half the observed speed. If the speed is much lower, there is no need to invoke exotic materials or shapes, only the requirement that the meteoroid be stronger compared to the background population of fireballs \citep{Borovicka2020}, but still within the range of tensile strength of stony meteorites. 

The factor of several difference between the infrasound source energy and USG lightcurve-derived energy may be due to a smaller luminous efficiency used by the USG website compared to the actual value of $\tau$. The USG efficiency was derived empirically from a subsample of USG fireballs where infrasound signals or meteorites were available to calibrate impactor size/energy \citep{Brown2002} at characteristic speeds of 15-18 km/s and peak brightness heights of $\sim$ 30km. For the lower altitude and probable higher speed of USG 2010108, a larger $\tau$ is likely, potentially explaining part of the difference \citep{Nemtchinov1997a}. A similar difference may arise if the effective blackbody temperature of the fireball was different than the USG assumed value of 6000K \citep{Ceplecha1998}.

Additionally, the low fragmentation height of the fireball could also lead to smaller fundamental acoustic periods than the average USG events used to produce period-yield relations for bolides \citep{Gi2017}. Since the blast radius for fragmentation events scales with ambient pressure as $P^{-1/3}$ \citep{McFadden2021}, energy release at lower heights will produce smaller (apparent) source energies using the average relations as periods will be smaller.

One interpretation of all the foregoing data could be that USG 20140108 was produced by an iron meteoroid. Irons constitute about 3.6\% of all cm-sized fireballs observed by the European Network \citep{Vojacek2020} and just over 4\% of all observed meteorite falls \citep{Greenwood2020}. However, the fraction of iron meteorite falls increases sharply with mass \citep{Bland2006}, with 25\% of all meteorite falls of between 200kg and 1000 kg being irons. A clear strength selection effect is evident; the proportional increase in irons as a function of mass underscores the higher strength of the latter compared to stony meteorites and their resistance to fragmentation.

\begin{figure*}[!ht]
    \centering
    \includegraphics[width=.99\textwidth]{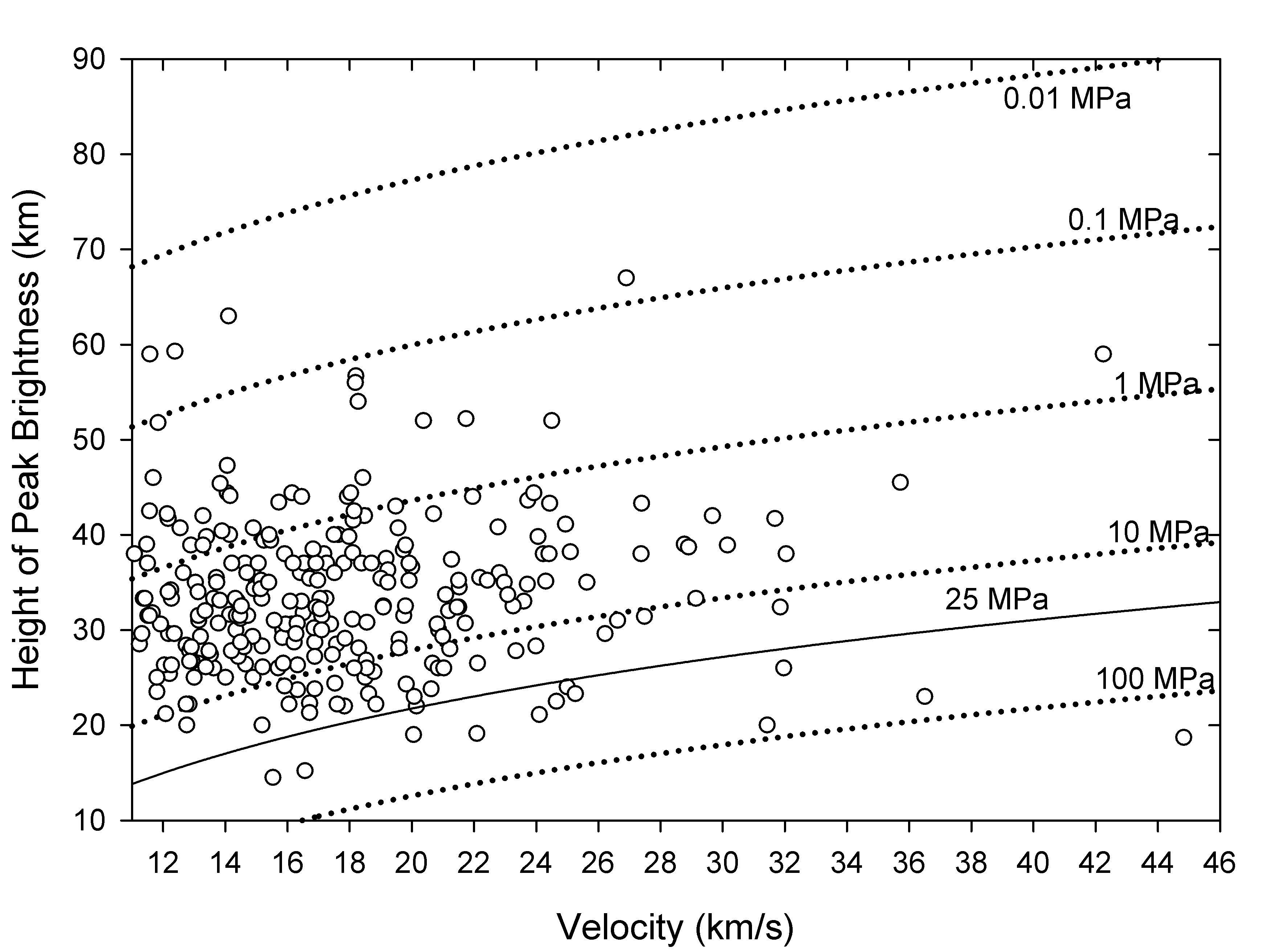}
    \caption{All USG fireballs where speeds and heights of peak brightness have been measured. The labelled curves are lines of equal dynamic pressure.}
    \label{fig:ht-peak}
\end{figure*}

\begin{figure*}[!ht]
    \centering
    \includegraphics[width=.99\textwidth]{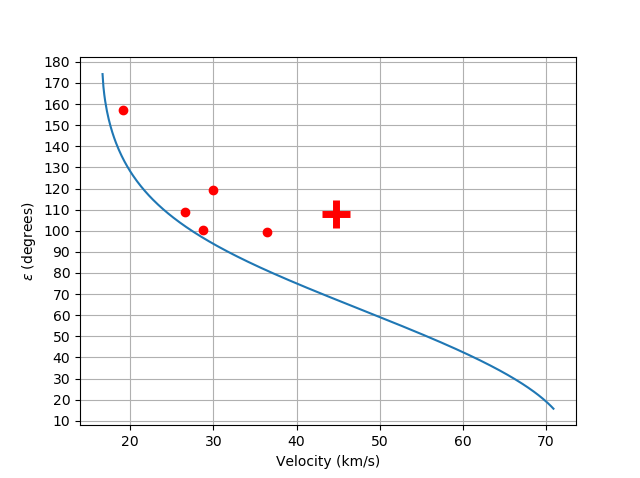}
    \caption{Radiant elongation ($\epsilon$) from the Earth's apex as a function of impact speed for all six USG events with nominally unbound orbits. USG 20140108 is the large red cross. The blue line delimits unbound orbits (to the right of the line) from those which are bound (to the left of the line).}
    \label{fig:hyper}
\end{figure*}

Assuming $\sim$5\% of all USG events are iron we would expect of order 15 to have been observed by USG and had speeds measured. Figure \ref{fig:ht-peak} shows the height of peak brightness for all USG fireballs as a function of speed overlaid on lines of equal dynamic pressure. The vast majority of all bolides detected by USG have their peak brightness at locations where the dynamic pressure is between 0.5 - 10 MPa.

While modest differences in height of peak brightness are expected for objects of similar composition given the random nature of cracks, populations separated by large peak height differences may reflect more significant physical property differences. For example, a population which reach peak brightness below 0.2-0.3 MPa dynamic pressure is noticeable in the top left of Figure \ref{fig:ht-peak} and indicative of unusually weak objects separated from the remaining population. 

Similarly, there are a dozen events where peak brightness is reached when dynamic pressure exceeds $\sim$25 MPa. These dozen events could be interpreted as being stronger on average than the majority of the USG population (if the speeds are correct) or perhaps as having internal components that are strong. This includes USG 20140108 (extreme point in the lower right hand corner) and also another fireball which occurred on 20170309 with a reported speed of 36.5 km/s and height of peak brightness of 23 km. This also happens to be the 2$^{nd}$ most unbound orbit in the USG dataset. The remaining four unbound USG fireballs (see Table \ref{tab:USG_prior}) are indistinguishable from the bound USG population in terms of their height of peak brightness, speed.

While an iron interpretation is attractive, the modelling in section \ref{sec:model} strongly argues against an iron object. This is because an iron would have much higher ablation coefficients than used in our modelling \citep{Ceplecha1998} and even at low speeds such high mass loss rates would cause the fireball to be visible much earlier than is observed. The object would also have to be much larger than modelled, further increasing the difficulty in matching the lightcurve. Moreover, irons do not typically show flares but rather have smooth light curves \citep{Vojacek2020} as would be physically expected. So, we can fit the light curve either with a low speed ($<$ 20 km/s) and a reasonable set of parameters or with a high speed combined with  low $\Gamma$A, low $\sigma$ and very high density/ strength. Such material cannot be iron but some dense material with high melting temperature. 

Finally, the one fireball observed by USG for which iron meteorites were recovered (Froslunda - see Table \ref{tab:USgcomp}) shows no distinct flares and at 17 km/s entry speed reached peak brightness at 23 km altitude. While this at the bottom of the dense USG population in Figure \ref{fig:ht-peak} it is still well above the deepest penetrators.

\section{Conclusion}\label{sec:con} 
Our modelling of the USG 20140108 fireball shows that we can fit the observed lightcurve and heights with a low speed ($<$20 km/s) using physical parameters consistent with a stony impactors as measured for other similar fireballs \citep{Borovicka2020}. The reported high speed can only be matched with an unusually low luminous efficiency, extremely aerodynamic (low drag) shape and high density/strength. We are able to rule out an iron composition for the latter as iron ablation proceeds via melting (rather than vaporization as is the case for stony meteoroids) which leads to high mass loss and a brightness too high to match the measured lightcurve and deep penetration of the fireball. 

The source energy derived from infrasound records is much smaller than the nominal USG energy. We suggest the most likely explanation is the lower than average fragmentaton height produces an artifically small blast cavity which translates into smaller periods and hence apparent source energies than the typical USG fireball population. 

Examination of the 17 USG fireballs having independent metric measurements show that the speed and radiant errors are large in many cases and vary widely in a non systematic manner. Higher speed events tend to be overestimated in speed, while radiant errors of order tens of degrees are common in the dataset.  
To resolve these contradictions access to the original metric data and an understanding of the measurement process is required. 
Ultimately this is the only means to conclude with confidence the hyperbolic or bound nature of USG 20140108. 

Appealing to Occam's razor we suggest that a significant error in the velocity vector is the most probable explanation for the apparent interstellar nature of USG 20140108 (i.e. it did not have an unbound orbit prior to Earth impact). This follows from our determination of the apparent USG uncertainties, and the fact that for any fixed time sampling observing system the relative meteor velocity error will increase with apparent speed \citep{vida2020}. 

Based on the foregoing analysis as USG 20140108 is at the very upper limit of speeds reported by USG systems, the required error in velocity is compatible with moving the orbit from an unbound to bound state. Moreover, the extraordinary physical properties of the associated meteoroid required to model the height and brightness behaviour of the event if it is high speed can instead be matched using properties similar to other chondritic fireballs presuming the measured speed and/or radiant are in error.

A similar conclusion is reached by both \citet{Vaubaillon2022} and \citet{ hajdukova23}.

\section{Acknowledgements} \label{sec:acknowledgments}
PGBs contribution was made possible by funding under NASA Meteoroid Environment Office Cooperative agreement 80NSSC21M0073 and by the Natural Sciences and Engineering Research Council of Canada (Grants no. RGPIN-2018-05659)as well as by the Canada Research Chairs Program. JB was supported by grant no.\ 19-26232X from the Czech Science Foundation.


\bibliography{References}{}
\bibliographystyle{aasjournal}



\appendix

\begin{sidewaystable}
    \centering
    \footnotesize
    \setlength\tabcolsep{4pt}
    \begin{longtable}{|l|l|l|l|l|l|l|l|l|l|l|l|l|l|l|l|l|}
    \caption{All observed USG fireballs with hyperbolic orbits. All angular elements are in degrees and J2000.0. Here Ht refers to height at peak brightness and Vel is apparent in atmosphere speed in km/s. The total radiated energy (J) as given on the USG website is also shown, together with each fireball's orbital elements. $\epsilon$ is the elongation of the apparent radiant from the apex of the Earth's way, V$_h$ is the heliocentric velocity of the bolide and $\alpha_g$ and $\delta_g$ are the geocentric radiant coordinates. \label{tab:USG_prior}} \\
    \hline
    Time (UT) & Lat & Long & Ht (km) & Vel & Rad Energy (J) & $q$ (AU) & a (AU) & e & inc & $\omega$ & $\Omega$ & $\epsilon$ & Vh &$\alpha_g$ & $\delta_g$ \\ \hline
    \endfirsthead
    \hline
    Time (UT) & Lat & Long & Ht (km) & Vel & Rad Energy (J) & $q$ (AU) & a (AU) & e & i & $\omega$ & $\Omega$ & $\epsilon$ & Vh & $\alpha_g$ & $\delta_g$ \\ \hline
    \endhead
\endhead
2014-01-08 17:05 & 1.3S & 147.6E & 18.7 & 44.8 & $3.1\times10^{10}$ & 0.7 & -0.5 & 2.4 & 9.7 & 57.7 & 108.2 & 108.1 & 61.0 & 88.5 & 13.3 \\ \hline
2017-03-09 4:16 & 40.5N & 18.0W & 23.0 & 36.5 & $4.0\times10^{11}$ & 0.7 & -1.3 & 1.5 & 23.9 & 239.8 & 348.5 & 99.4 & 49.8 & 170.2 & 34.3 \\ \hline
2022-07-28 1:36 & 6.0S & 86.9W & 37.5 & 29.9 & $2.5\times10^{11}$ & 0.9 & -1.9 & 1.5 & 23.4 & 221.0 & 124.7 & 119.4 & 47.0 & 276.1 & 14.8 \\ \hline
2009-04-10 18:42 & 44.7S & 25.7E & 32.4 & 19.1 & $2.7\times10^{11}$ & 1.0 & -3.6 & 1.3 & 6.2 & 357.7 & 200.9 & 157.2 & 44.9 & 108.0 & 4.2 \\ \hline
2021-05-06 5:54 & 34.7S & 141.0E & 31.0 & 26.6 & $2.1\times10^{10}$ & 0.7 & -6.6 & 1.1 & 5.7 & 298.2 & 225.6 & 108.8 & 43.5 & 61.9 & 12.1 \\ \hline
2015-02-17 13:19 & 8.0S & 11.2W & 39.0 & 28.8 & $3.3\times10^{10}$ & 0.6 & -10.5 & 1.1 & 0.4 & 286.9 & 148.5 & 100.5 & 43.4 & 339.3 & -9.3 \\ \hline
    \end{longtable}
\end{sidewaystable}

\section{Infrasound Signals} \label{sec:infra}

\begin{figure*}[!ht]
    \centering
    \includegraphics[width=.49\textwidth]{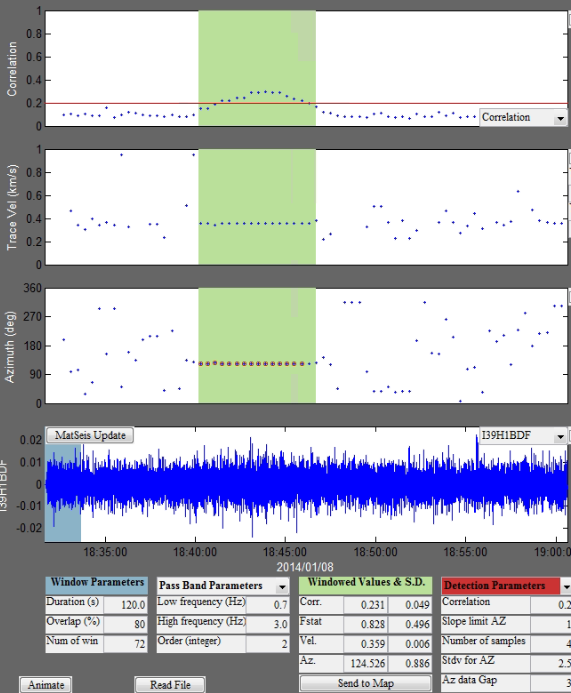}
    \caption{Beamformed array arrivals of the infrasound from USG 20140108 as detected at the I39PW array in Palau some 1750 km from the fireball. Window lengths of two minutes across the seven element array were stepped with 80\% overlap to find the best beam direction per window. The top plot shows the signal correlation in each window - here the coherent airwave is visible for values of the coefficient above 0.2. The second plot shows the apparent speed of the signal across the array while the third shows the backazmiuth of the signal in each time window. The bottom plot is the pressure signal at array element 1 bandpassed from 0.7-3 Hz.  The signal is highlighted in green.}
    \label{fig:infra-I39}
\end{figure*}

Figure \ref{fig:infra-I39} shows an example of the beamformed signal as detected at the closest station, I39PW.

\begin{figure*}[!ht]
    \centering
    \includegraphics[width=.49\textwidth]{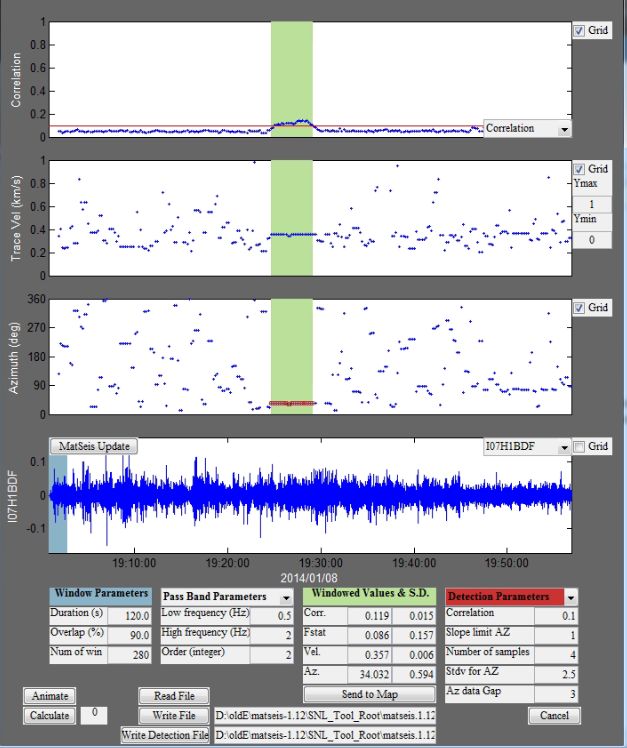}
    \caption{Beamformed array arrivals of the infrasound from USG 20140108 as detected at the I07AU array in Australia at a range of 2525 km. Window lengths of two minutes across the seven element array were stepped with 90\% overlap to find the best beam direction per window. The top plot shows the signal correlation in each window - here the coherent airwave is visible for values of the coefficient above 0.1. The second plot shows the apparent speed of the signal across the array while the third shows the backazmiuth of the signal in each time window. The bottom plot is the pressure signal at array element 1 bandpassed from 0.5-2 Hz.  The signal is highlighted in green.}
    \label{fig:infra-I07AU.jpg}
\end{figure*}

\begin{figure*}[!ht]
    \centering
    \includegraphics[width=.49\textwidth]{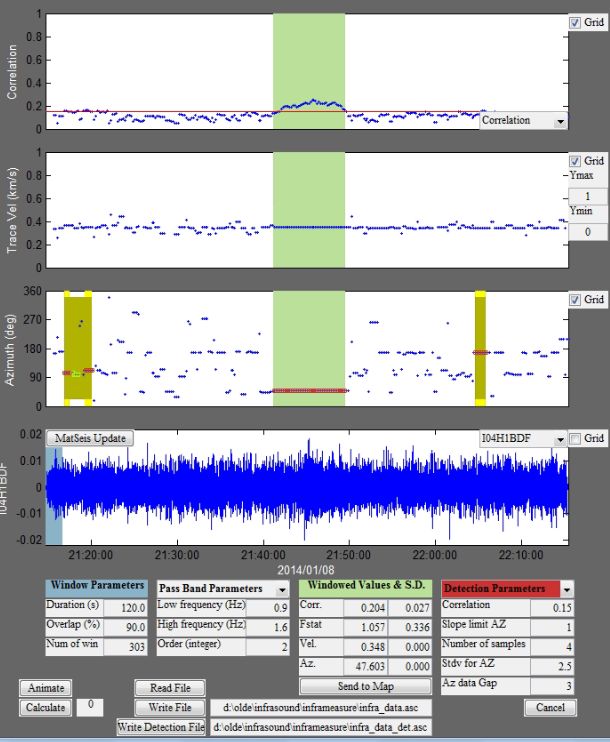}
    \caption{Beamformed array arrivals of the infrasound from USG 20140108 as detected at the I04AU array in Australia at a range of 4820 km. Window lengths of two minutes across the seven element array were stepped with 90\% overlap to find the best beam direction per window. The top plot shows the signal correlation in each window - here the coherent airwave is visible for values of the coefficient above 0.15. The second plot shows the apparent speed of the signal across the array while the third shows the backazmiuth of the signal in each time window. The bottom plot is the pressure signal at array element 1 bandpassed from 0.9-1.6 Hz.  The signal is highlighted in green.}
    \label{fig:infra-I04AU.jpg}
\end{figure*}


\end{document}